\documentclass[aps,pra,twocolumn]{revtex4}
\usepackage[latin1]{inputenc}
\usepackage{epsfig}
\usepackage{hhline}
\usepackage{amsmath,amssymb,amsfonts}
\usepackage{float}
\usepackage{subfig}
\usepackage{xspace}
\usepackage{xcolor}
\usepackage{ulem}
\usepackage{dcolumn}

\newcolumntype{.}{D{.}{.}{-1}}

\begin{document}

\captionsetup{justification=centerlast}

\title{\vspace{2cm}\textbf{Routine calculation of {\it ab initio} melting curves: application to aluminum}}
\author{Grégory Robert, Philippe Legrand, Philippe Arnault, Nicolas Desbiens, Jean Clérouin}
\affiliation{CEA, DAM, DIF, F-91297 Arpajon, France}

\begin{abstract}
We present a simple, fast, and reliable method to compute the melting curves of materials with {\it ab initio} molecular dynamics. It is based on the two-phase thermodynamic model of [Lin {\it et al.}, J. Chem. Phys. \textbf{119}, 11792 (2003)] and its improved version given by [Desjarlais, Phys. Rev. E, \textbf{88}, 062145 (2013)]. In this model, the velocity autocorrelation function is utilized to calculate the contribution of the nuclei motion to the entropy of the solid and liquid phases. It is then possible to find the thermodynamic conditions of equal Gibbs free energy between these phases, defining the melting curve. The first benchmark on the face-centered cubic melting curve of aluminum from 0 to 300\,GPa demonstrates how to obtain an accuracy of 5-10\%, comparable to the most sophisticated methods, for a much lower computational cost.
\end{abstract}

\medskip
\maketitle

\section{Introduction}

The knowledge of the melting curve of materials is of great importance for condensed matter physics. At high pressures, it is of high interest to planetary sciences \cite{Baraffe2010}, and other fields of technological interest, such as inertial confinement fusion \cite{Lindl2004}. Its determination relies mostly on calculations, such as {\it ab initio} simulations, validated on the restricted domain of pressure and temperature accessible experimentally. Although well-defined by the condition of equal Gibbs free energy between both solid and liquid phases, the thermodynamic conditions of melting remain challenging to predict theoretically \cite{bouchet} and to observe experimentally at high pressures, even leading sometimes to controversial results \cite{Burakovsky2010, desjarlais}. Whereas large numerical efforts and/or experimental investments are required to get accurate melting curves for selected chemical elements, there is presently no routine procedure to get reliable, and accurate enough, calculations of melting curves for materials and applications that do not deserve the burden of more involved treatments. The present study demonstrates that a rapid and reliable calculation of melting curves at high pressures can be obtained thanks to Desjarlais's recent proposal of a first-principles calculation of entropy for liquid metals \cite{desjarlais}.
\\\indent
The reference method for melting curve calculation is the {\it thermodynamic integration} \cite{vocadlo}. However, this method gives rise to an overwhelming computational cost, since the Helmholtz free energy, and the liquid entropy in particular, is obtained from the internal energy of many {\it ab initio} simulations sliding continuously from the actual system to a reference system with known properties. Other less expensive {\it ab initio} molecular dynamics methods have also faced problems such as too large computational times, too small cell sizes, and too sensitive monitoring of the simulations \cite{bouchet}. The most used methods, to our knowledge, are the followings, in order of increasing sophistication.\\
{\it Heat Until it Melts (HUM):} A solid phase is heated along an isochore up to melting. This simple one-phase approach has a very low numerical cost. Unfortunately, the melting temperature is, in some cases, overestimated by 20 to 30\% since the homogeneously heated solid phase stays metastable above the true melting temperature. The obtained temperature corresponds rather at best to the highest temperature $T_h$ which can be reached by a superheated solid.\\
{\it Hysteresis} \cite{luo}: HUM is supplemented by the reverse process, a liquid phase is cooled down to the lowest temperature $T_l$, which can be reached by a supercooled liquid. An empirical average of the two temperatures, $T_h$ and $T_l$, based on the homogeneous nucleation theory \cite{Kelton1991}, gives an estimate of the true melting curve. However, the monitoring of the heating or cooling of the metastable phases to reach the very values of $T_l$ and $T_h$ can become tricky in some cases.\\
{\it Z method} \cite{belonoshko,belonoshko2}: A simulation in the microcanonical ensemble is started from a solid phase where an initial energy deposit is given. The system then evolves along an isochore, which can be decomposed into a solid part, a superheated solid part (that extends into the liquid region above the melting temperature) and a transition from the highest superheated solid state down to the melting curve. This method, which gives an upper bound to the melting temperature, can become very time consuming at high pressures.\\
{\it Two-phases approach} \cite{alfe,belonoshko3}: A solid slab is in contact with a liquid slab at a given temperature. The interface is the seed to the phase transition towards the liquid above the melting temperature, or towards the solid below it. The melting temperature is bracketed progressively by dichotomy. 
This method needs large cells and long simulations to properly describe the interface and to follow its overall evolution. Although known to work well with classical potentials, applying this technique on {\it ab initio} simulations often outruns computational capabilities.

In this paper, we shall propose an alternative method, whose accuracy is comparable to the most sophisticated methods, for a much lower computational cost. It utilizes Desjarlais's recent proposal \cite{desjarlais} for the calculation of liquid entropy. Desjarlais has enhanced the two-phase thermodynamic (2PT) model of Lin {\it et al.} \cite{lin}. In the 2PT model, the velocity autocorrelation function (VACF) is decomposed into gas-like (diffusive) and solid-like (vibratory) components. The Fourier transform of the solid-like component gives the vibrational density of states (DOS) and its contribution to the entropy, assuming each mode is a harmonic oscillator. The gas-like contribution is given by the hard sphere (HS) entropy. Using a memory function (MF) formalism, Desjarlais \cite{desjarlais} has corrected the short time (high frequency) non-physical behavior of the HS VACF, used originally, while preserving the long time (low frequency) correlation time scale of Lin {\it et al.}'s model \cite{lin}. This has resulted in an improved accuracy of the calculated entropies for softer interatomic potentials. Desjarlais has used the 2PT-MF model only for the liquid entropy calculation to demonstrate its accuracy whereas his solid entropy calculation uses a combination of quasiharmonic phonon calculations followed by a thermodynamic integration in temperature along isochores to obtain the anharmonic contributions. He has validated the 2PT-MF model against experimental data along the ambient isobar for liquid sodium, aluminum, tin, gallium and iron. 
\\\indent
Our application of the 2PT-MF model to the determination of melting curves is different from Desjarlais's first example in several respects. We have utilized the 2PT-MF model for both solid and liquid phases with the purpose of defining the more simple and straightforward method. We have also stuck to the usual simulation conditions (equilibration phase, time step,...) to highlight the performances of the method. As a consequence, we did not optimize the time step of the simulation nor its duration with respect to the VACF accuracy needed in the 2PT-MF model. Finding the most appropriate input parameters for the determination of 2PT-MF melting curves would necessitate an extensive study comprising different chemical elements, that we leave for future work. 

In the following section, the 2PT-MF model is described for completeness. Then the third section details the methodology we propose and, as an example, we apply this method to the face-centered cubic (fcc) melting curve of aluminum extensively studied both experimentally \cite{Boehler1997, Hanstrom2000, Shaner1984} and theoretically
\cite{moriarty,mei,morris,dewijs,jesson,alfe,bouchet} and make comparisons with published results in the fourth section.

\section{2PT-MF model in a nutshell}

Since the pressure $P = P(V,T)$ and the internal energy $E = E(V,T)$ are direct outputs of the molecular dynamics codes, and
\begin{equation}
G = E -T\,S + P\,V,
\end{equation}
the only missing quantity is the entropy, $S = S(V,T)$, which includes an electronic contribution, $S^e$, part of the free energy functional minimized in {\it ab initio} simulations and a contribution, $S^i$, due to the nuclei motion. This last contribution is evaluated here within the 2PT-MF model \cite{desjarlais}. First, $S^i$ is split into gas-like ($S^i_g$) and solid-like ($S^i_s$) parts
\begin{equation}
S^i = S^i_g + S^i_s.
\end{equation}
For $N$ atoms, a fraction $f_g$ of the $3N$ degrees of freedom represents the gas-like contribution, approximated by the HS entropy
\begin{subequations}
\begin{equation}
S^i_g = 3 N k f_g W_g,
\end{equation}
with Sackur--Tetrode ideal gas (IG) and excess ($x$) contributions
\begin{equation}
W_g = W_\text{IG} + W_x,
\end{equation}
where
\begin{equation}
3\, W_\text{IG} = \frac{5}{2}+\ln\left[\left(\dfrac{2 \pi m k T}{h^2}\right)^{3/2} \dfrac{V}{f_g N}\right],
\end{equation}
and
\begin{equation}
3\, W_x = \ln\left[\dfrac{1+\gamma+\gamma^2-\gamma^3}{(1-\gamma)^3}\right] + \dfrac{\gamma (3 \gamma -4)}{(1-\gamma)^2}.
\end{equation}
In the former equations, $k$ is Boltzmann's constant, $h$ Planck's constant, $m$ the atomic mass, and $\gamma$ the HS packing fraction, solution to \cite{lin}
\begin{equation}
\dfrac{2\,(1-\gamma)^3}{2-\gamma}-\gamma^{2/5} \Delta^{3/5} = 0,
\end{equation}
with
\begin{equation}
\Delta = \dfrac{8}{3} \left(\dfrac{6}{\pi}\right)^{2/3}\,D\,\sqrt{\dfrac{\pi m}{k T}}\,\left(\dfrac{N}{V}\right)^{1/3},
\end{equation}
\end{subequations}
where $D$ is the diffusion coefficient obtained from the VACF, $Z(t)$, by
\begin{subequations}
\begin{equation}
D = \int_0^\infty Z(t) dt,
\end{equation}
if we define the VACF by
\begin{equation}
Z(t) = \dfrac{1}{3 N} \sum_{i=1}^N\,\lim_{T\to \infty} \dfrac{1}{T} \int_0^T \vec u_i(t+\tau)\cdot \vec u_i(\tau) d\tau,
\end{equation}
\end{subequations}
where $\vec u_i(t)$ is the velocity of atom $i$ at time $t$.

\begin{figure}
 \subfloat{ \includegraphics[scale=0.216]{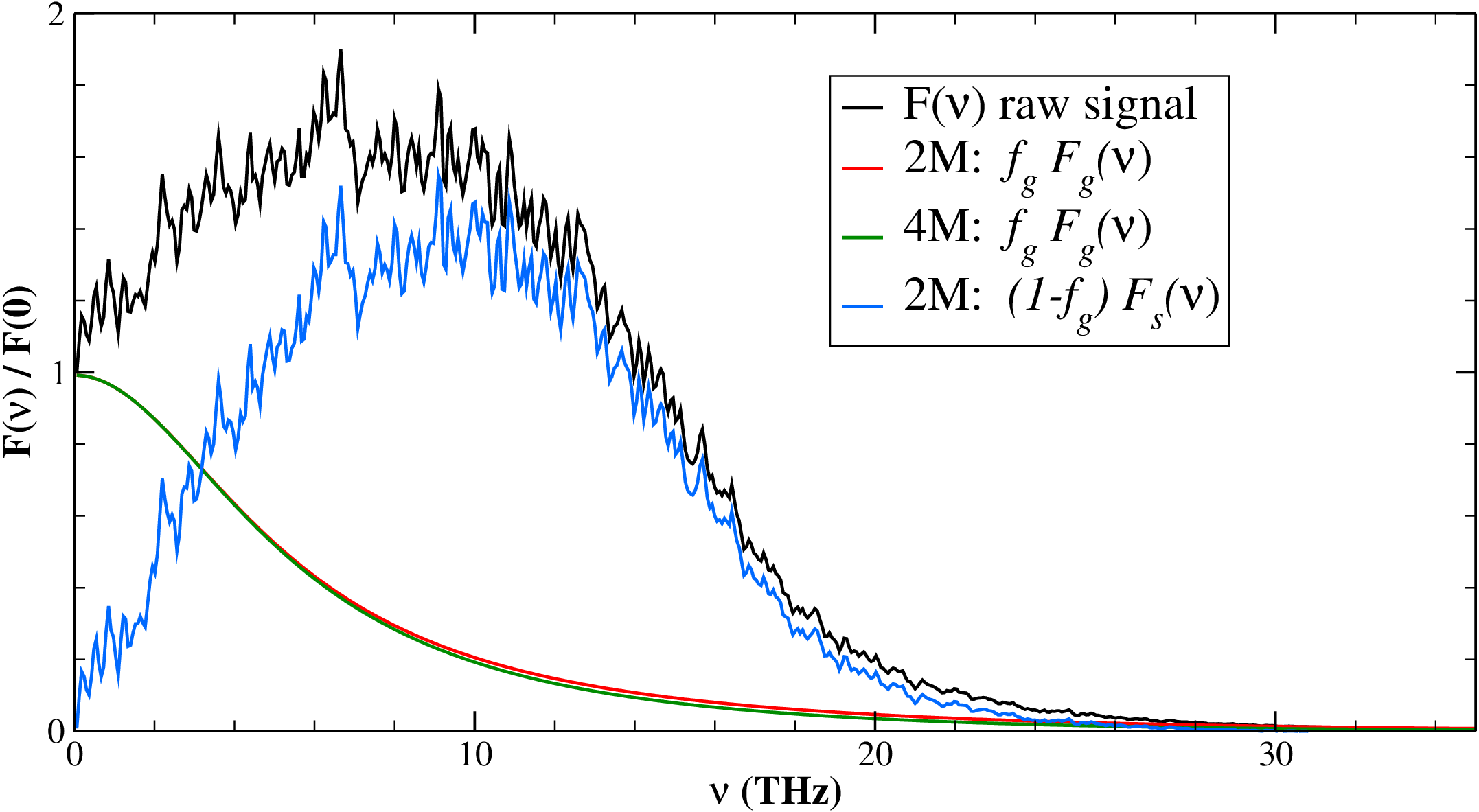} }\\
 \subfloat{ \includegraphics[scale=0.216]{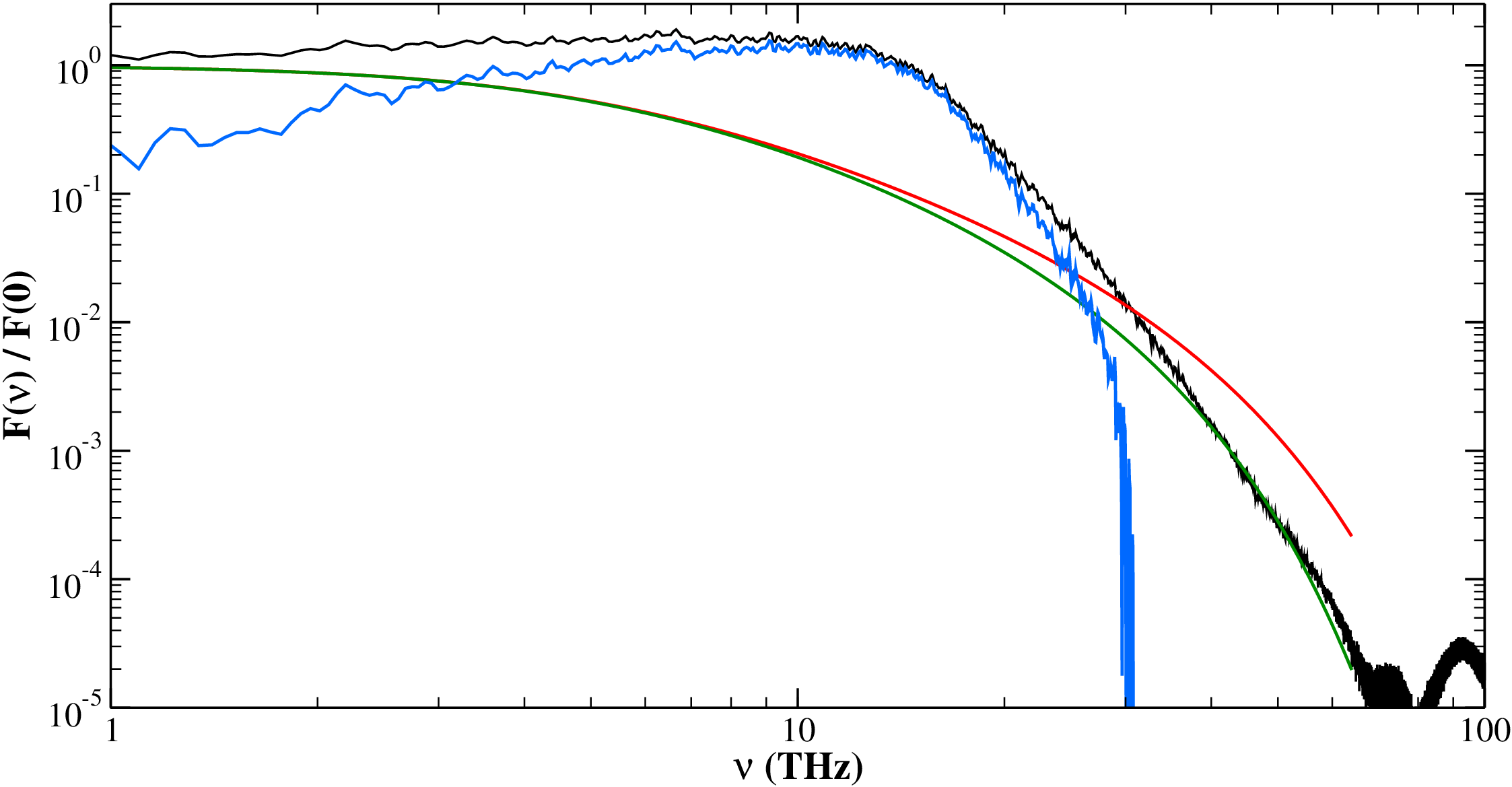} }
 \caption{(Color online) Fourier transform $F(\nu)$ of the VACF of aluminum at 4000 K and 0.3125 cm$^3$\,g$^{-1}$ and the gas-like and solid-like contributions in linear-linear plot (top) and in log-log plot (bottom). Black line is the raw $F(\nu)$ signal, red (green) line is the 2M (4M) gas-like contribution and the blue line is the solid-like DOS deduced from the 2M method.}
 \label{fig:DOS}
 \end{figure}

The solid-like contribution to the entropy, $S^i_s$, is obtained from
\begin{subequations}
\begin{equation}
S^i_s = N k\, (1 - f_g) \int_0^\infty F_s(\nu) W_s(\nu) d\nu,
\end{equation}
where $F_s$ is the vibrational DOS (normalized to 3), and $W_s$ is the contribution of each harmonic mode
\begin{equation}
W_s(\nu) = \dfrac{h \nu/k T}{e^{h \nu/k T} - 1} - \ln\left[1-e^{-h \nu / k T}\right].
\end{equation}
The DOS $F_s$ is obtained subtracting from the Fourier transform $F$ of the VACF a properly defined gas-like contribution $F_g$ (Fig.\,\ref{fig:DOS})
\begin{equation}
(1 - f_g)\, F_s(\nu) = F(\nu) - f_g\, F_g(\nu),
\end{equation}
with
\begin{equation}
F(\nu) = \dfrac{12 m}{k T} \int_0^\infty Z(t) \cos(2 \pi \nu t)\, dt.
\end{equation}
Desjarlais \cite{desjarlais} has chosen a Gaussian memory function $K_g$ to represent the gas-like VACF
\begin{equation}
K_g(t) = A_g\,e^{-B_g t^2},
\end{equation}
the Laplace transform of which is
\begin{equation}
\hat K_g(z) = A_g\, \sqrt{\dfrac{\pi}{4 B_g}}\, e^{z^2/4 B_g}\, \mathtt{Erfc}\left[\dfrac{z}{2 \sqrt{B_g}}\right],
\end{equation}
where Erfc is the complementary error function. The gas-like contribution $F_g$ is then given by
\begin{equation}
\label{Fnu}
F_g(\nu) = 6 \left[ \dfrac{1}{\hat K_g(i 2 \pi \nu) + i 2 \pi \nu} + \dfrac{1}{\hat K_g(-i 2 \pi \nu) - i 2 \pi \nu}\right].
\end{equation}
A python script is provided in Appendix to compute the function $F_g(\nu)$ easily.
\end{subequations}
\\\indent
To complete the 2PT-MF model, three independent relations are needed to compute the three unknown parameters: $f_g$, $A_g$, and $B_g$. First, the zero-frequency limit gives
\begin{subequations}
\begin{equation}
\label{zeroF}
A_g = 2 f_g\,\sqrt{\dfrac{B_g}{\pi}}\,\dfrac{k T}{m D}.
\end{equation}
Second, the low-frequency (long-time) behavior, given by the HS analysis of Lin {\it et al.} \cite{lin}, requires
\begin{equation}
\label{lowF}
\dfrac{4 B_g}{A_g} = 2+\sqrt{\pi\,(1+4 B_g/\alpha^2)},
\end{equation}
with
\begin{equation}
\alpha = \dfrac{k T}{m D} \gamma^{2/5} \Delta^{3/5}.
\end{equation}
The third relation, provided by Desjarlais \cite{desjarlais}, utilizes the high-frequency (short-time) behavior of $F(\nu)$ assuming that the VACF $Z(t)$ is split into gas-like and solid-like components, each represented by its own Gaussian memory function. Actually, the memory function is a constant for the solid, $K_s = A_s$, since its diffusion vanishes. Then, the even frequency moments $M_{2n}$ of $F(\nu)$,
\begin{equation}
M_{2n} = \dfrac{1}{3}\,\int_0^\infty (2 \pi \nu)^{2 n} F(\nu)\, d\nu,
\end{equation}
are related to the memory function parameters by
\begin{align}
\label{twoM}
&M_2 = (1-f_g) A_s + f_g A_g,\\\nonumber
&M_4 = (1-f_g) A_s^2 + f_g (A_g^2 + 2 A_g B_g),
\end{align}
which gives a closed system of equations to solve with Eqs.\,\eqref{zeroF} and \eqref{lowF}, referred to in the following as the "two-moments" (2M) model. 
\\\indent
A "four-moments" (4M) model is provided representing the solid-like component by two Gaussian memory functions, leading to the following system
\begin{align}
\label{fourM}
M_2 = & f_1 A_1  + f_2 A_2 + f_g A_g,\\\nonumber
M_4 = & f_1 A_1^2 + f_2 A_2^2 + f_g (A_g^2 + 2 A_g B_g) \\\nonumber
M_6 = & f_1 A_1^3 + f_2 A_2^3 + f_g (A_g^3 + 4 A_g^2 B_g + 12 A_g B_g^2) \\\nonumber
M_8 = & f_1 A_1^4 + f_2 A_2^4 \\\nonumber
   + & f_g (A_g^4 + 6 A_g^3 B_g + 28 A_g^2 B_g^2 + 120 A_g B_g^3)  \\\nonumber
f_1 + &  f_2 + f_g = 1.
\end{align}
\end{subequations}

\begin{figure*}[t!]
\includegraphics[scale=0.35]{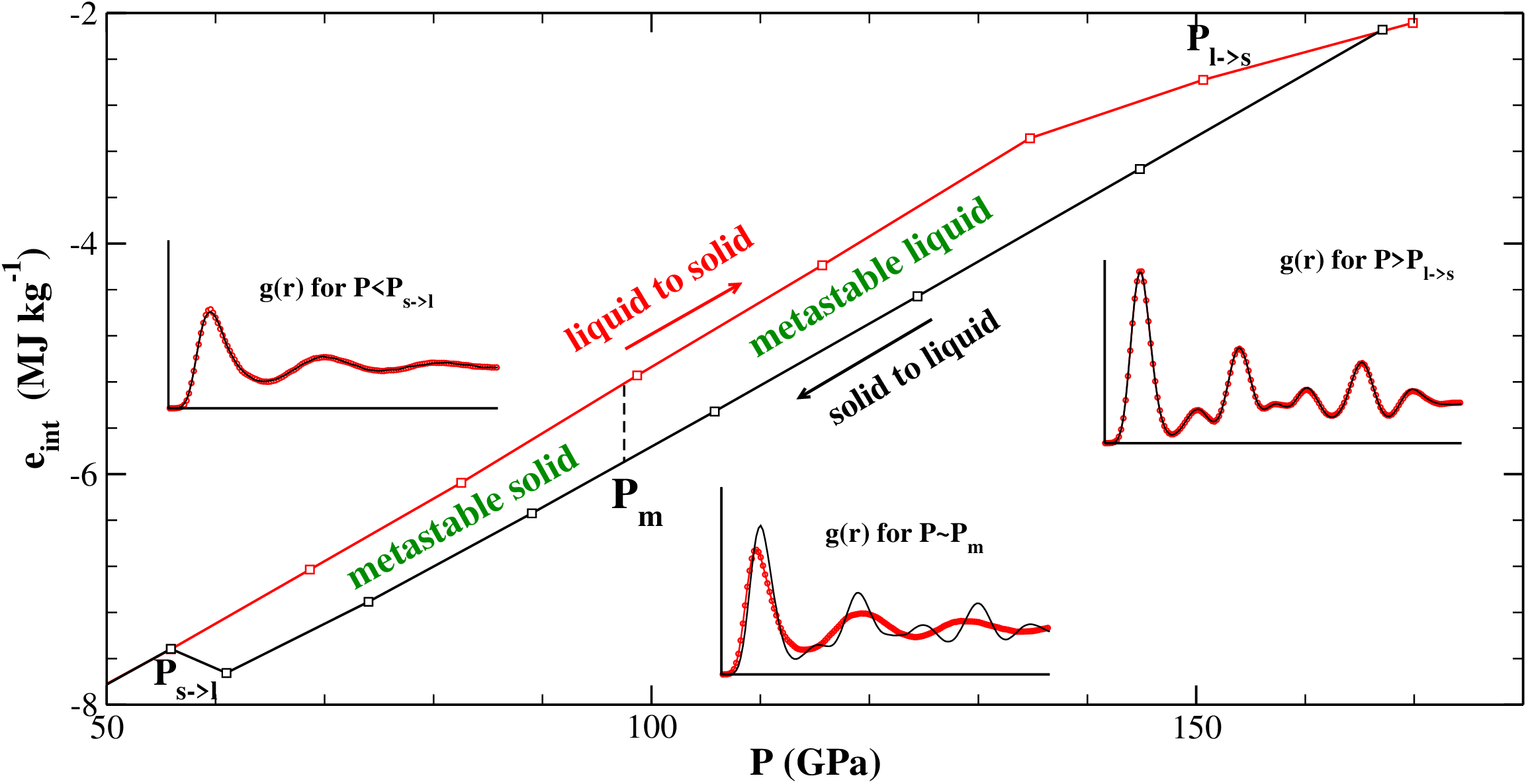}
\caption{\label{fig:schema}(Color online) Internal energy for the melting (black) and solidification (red) branches along the 4000\,K isotherm of aluminum. $P_m$ indicates the melting pressure, $P_{l\to s}$ (respectively $P_{s\to l}$) is the highest (lowest) pressure for which liquid (fcc solid) phase is metastable. Insets display typical radial distribution functions g(r) at $P<P_{s\to l}$, $P\thicksim P_m$ and $P>P_{l\to s}$. Colors of the $g(r)$ correspond to the melting or solidification branch.}
\end{figure*}

In this last step, some caution is in order since the high-order moments of $F(\nu)$ probe the high frequency tail of $F(\nu)$ which is dominated by noise when the time step is not sufficiently small and the sampling not extensive enough. We have found that a truncation of $F(\nu)$ at high frequency, five decades below its highest value (around 65 THz in Fig.\,\ref{fig:DOS}), allows generally to match the high-frequency tail with the gas-like component. 

\section{Methodology}

\begin{figure}[!hb]
\subfloat{  \includegraphics[scale=0.225]{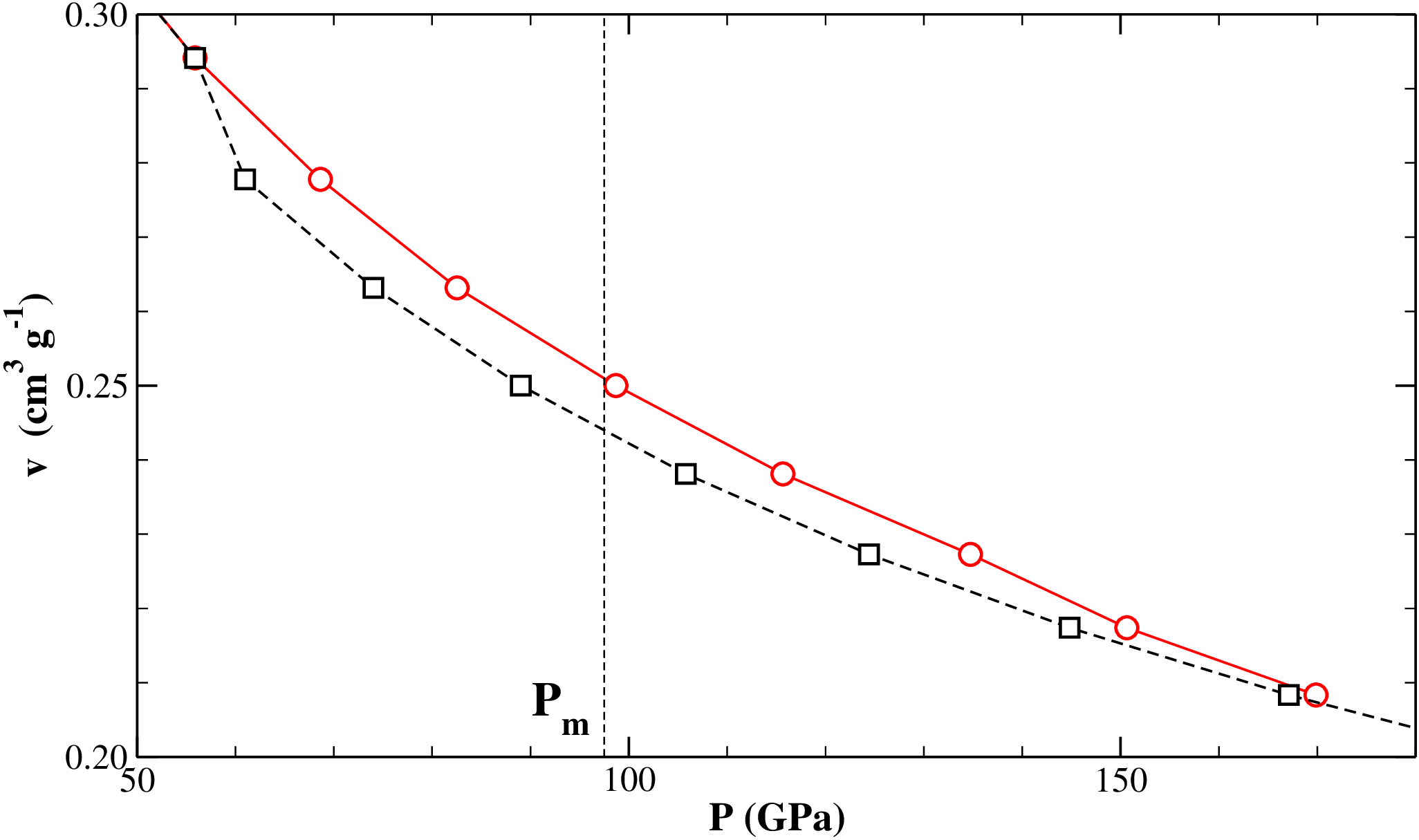} } \\
\subfloat{  \includegraphics[scale=0.225]{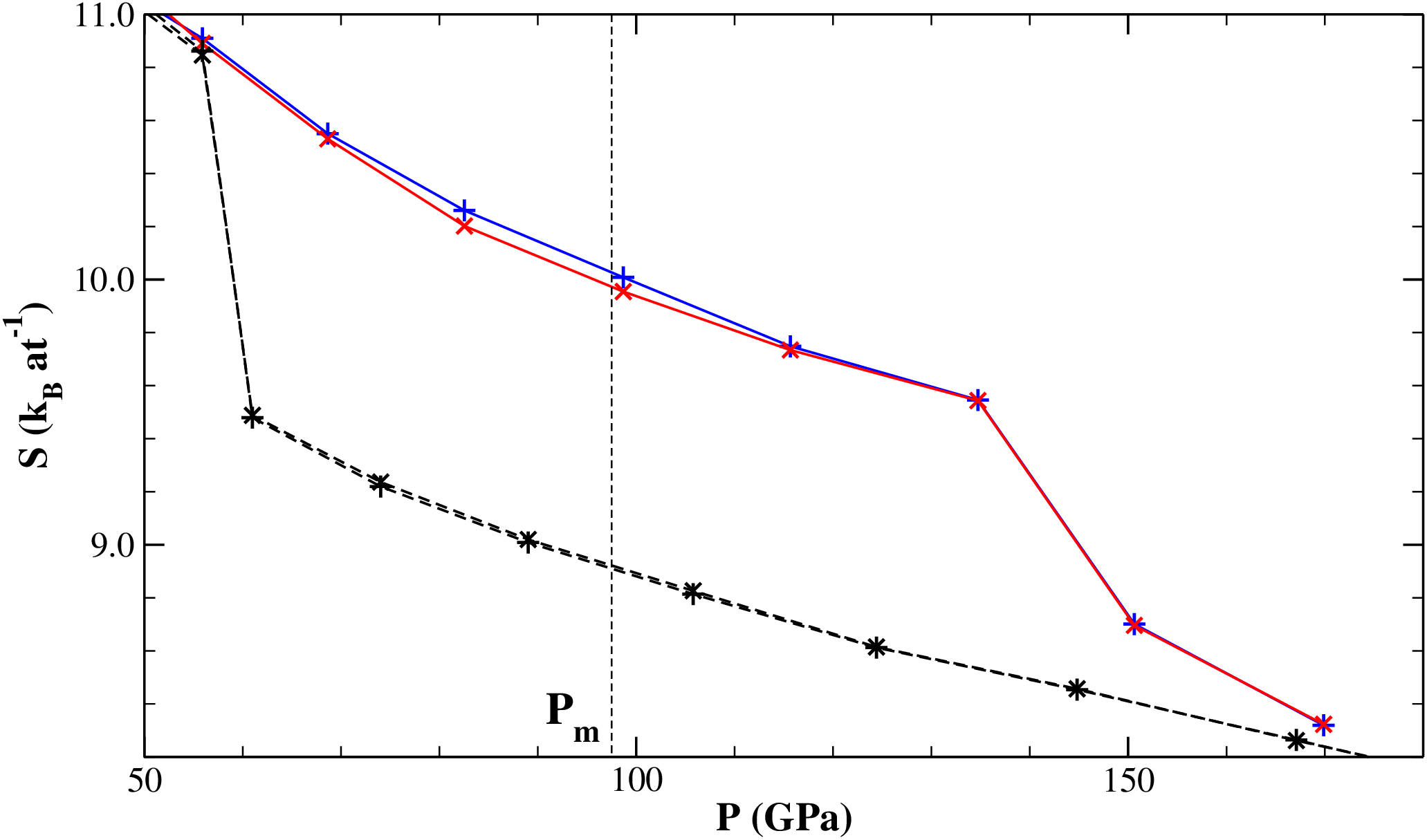} }
\caption{(Color online) Volume (top) and entropy (bottom) for the melting (black) and solidification (red) branches along the 4000 K isotherm of aluminum. $P_m$ is the melting pressure. For entropy, plus (cross) symbols correspond to 2M (4M) method.}
 \label{fig:schemab}
\end{figure}

From a practical viewpoint, since the {\it ab initio} molecular dynamics codes start usually from solid-state initial configurations, the simulations along the isotherms are conducted in a very similar way than in the hysteresis method \cite{luo}. Nevertheless, at contrast to the hysteresis method, the precise values of the limiting pressures, $P_{s\to l}$ and $P_{l\to s}$, of the solid and liquid metastable phases is not required. For each temperature $T_m$, a melting branch is computed starting from a perfect solid phase decreasing the density. For the corresponding solidification branch, one has to take care that the starting point is in a "true" liquid phase. To this end, we have monitored the radial distribution functions (RDF), whose typical shapes are reported in Fig.\,\ref{fig:schema} which represents the specific internal energy $e$ of liquid and solid phases of aluminum as a function of pressure $P$ along the 4000\,K isotherm. In the metastable zone between $P_{s\to l}$ and $P_{l\to s}$, we have checked that the RDF remains typical of the reference state (especially for the liquid-to-solid branch). The internal energy change $\Delta e$ from solid to liquid is almost constant along the range of pressure comprising both stable and metastable phases. This is also verified for the corresponding changes of specific entropy $\Delta s$ and volume $\Delta v$, as shown in Fig.\,\ref{fig:schemab}. This leads, in particular, to small variations of the compressibility at melting. A first estimate of the melting pressure $P_m$ for the temperature $T_m$ can therefore be given by the thermodynamic relation
\begin{equation}
\Delta e(T_m) = T_m\Delta s(T_m) - P_m \Delta v(T_m)\label{eq1}.
\end{equation}

\begin{figure}
\includegraphics[scale=0.205]{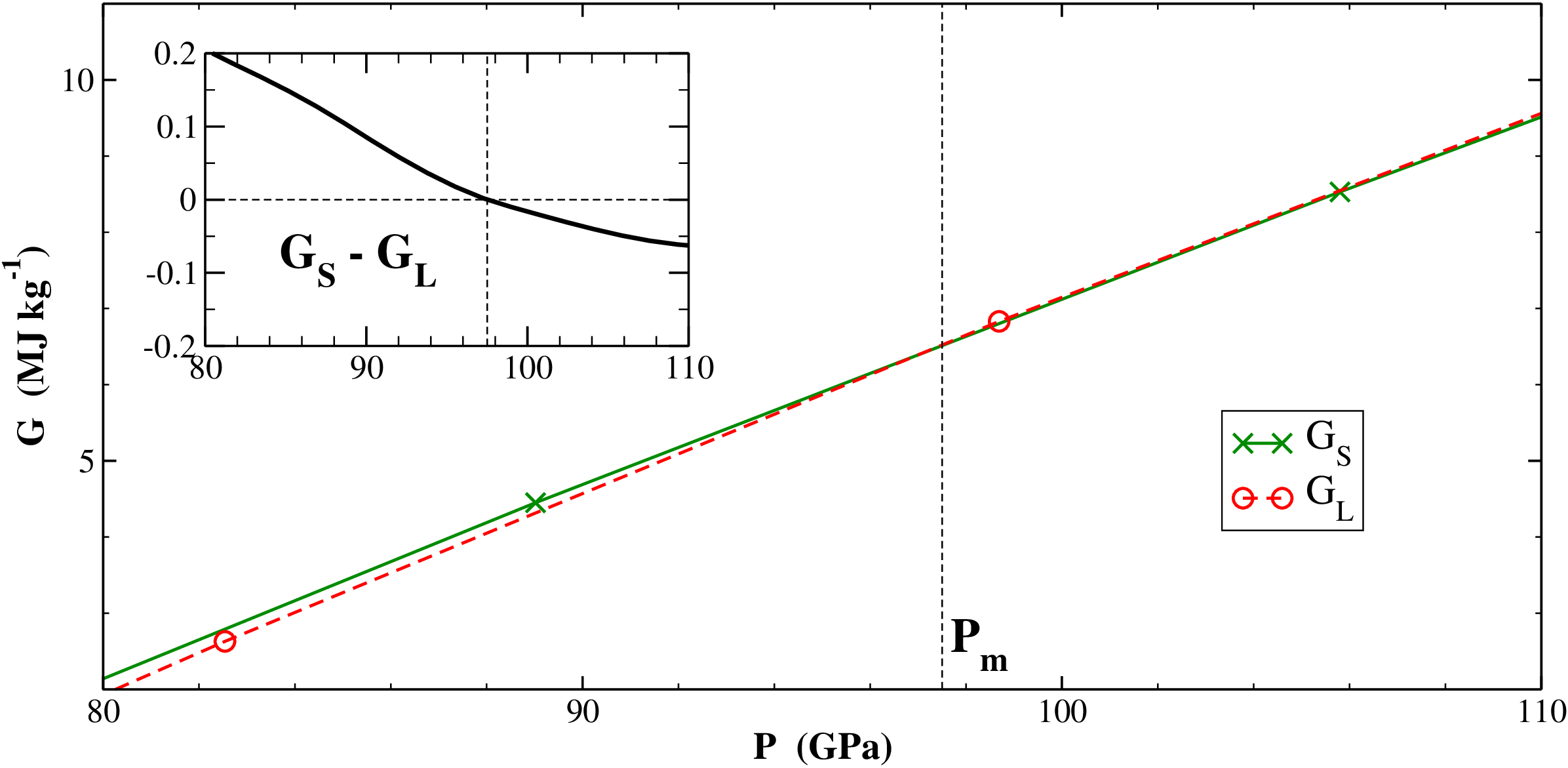}
\caption{(Color online) Gibbs free energies of the liquid and solid phases for the 4000 K isotherm of aluminum with the 4M method. Difference of Gibbs free energies are displayed in the inset.}
 \label{fig:schemac}
\end{figure}

More generally, even in the cases where the changes $\Delta e$, $\Delta s$ and $\Delta v$ are not constant in the metastable region, the intersection of the Gibbs free energies, $G_S$ and $G_L$, of the solid and liquid phases as functions of pressure $P$ along the $T_m$ isotherm (Fig.\,\ref{fig:schemac}), gives the melting pressure $P_m$ and the corresponding changes, $\Delta e(P_m)$, $\Delta v(P_m)$ and $\Delta s(P_m)$.

The Clapeyron's relation provides then the derivative of the melting temperature with respect to the melting pressure:
\begin{equation}
\dfrac{dT_m}{dP_m} = \dfrac{\Delta v(P_m)}{\Delta s(P_m)}.\label{clapeyron}
\end{equation}

\section{fcc melting curve of aluminum}

We consider the fcc melting curve of aluminum as a benchmark since many different theoretical calculations have been published \cite{moriarty,mei,morris,dewijs,jesson,alfe,bouchet} and there are experimental results at low and medium pressures \cite{Shaner1984,Boehler1997,Hanstrom2000}.

\begin{figure}[!t]
 \subfloat{  \includegraphics[scale=0.214]{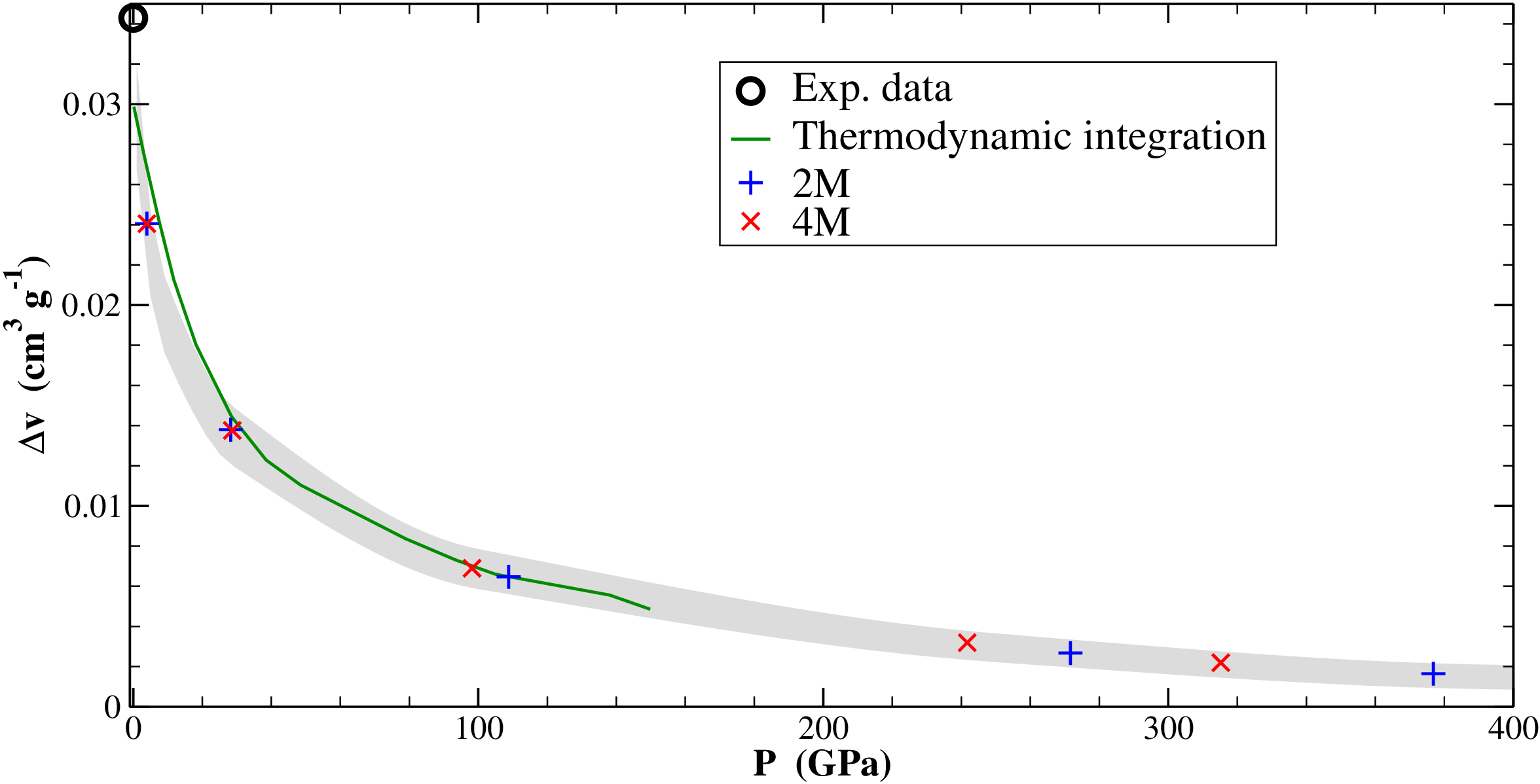}}\\
 \subfloat{  \includegraphics[scale=0.214]{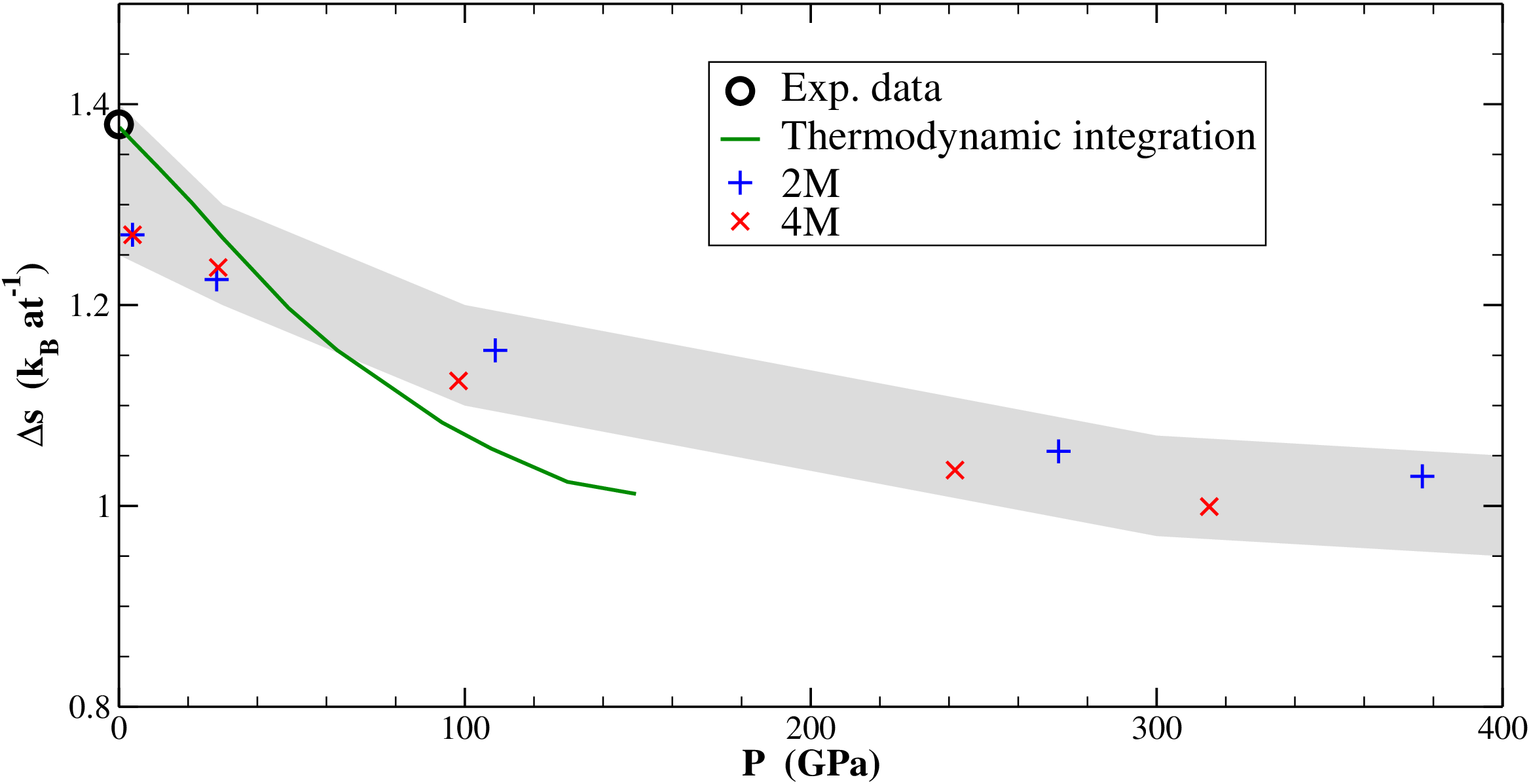}}
 \caption{(Color online) Volume (top) and entropy (bottom) changes at melting. Green lines correspond to the reference results of Vo{\v c}adlo and Alf\`e \cite{vocadlo}, open black circles are experimental values at ambient conditions \cite{cannon,chase}. Plus (cross) symbols stand for 2M (4M) method. Grey areas represent the estimated uncertainties.}
 \label{fig:Al_delta}
 \end{figure}

\begin{figure*}
\includegraphics[scale=0.4]{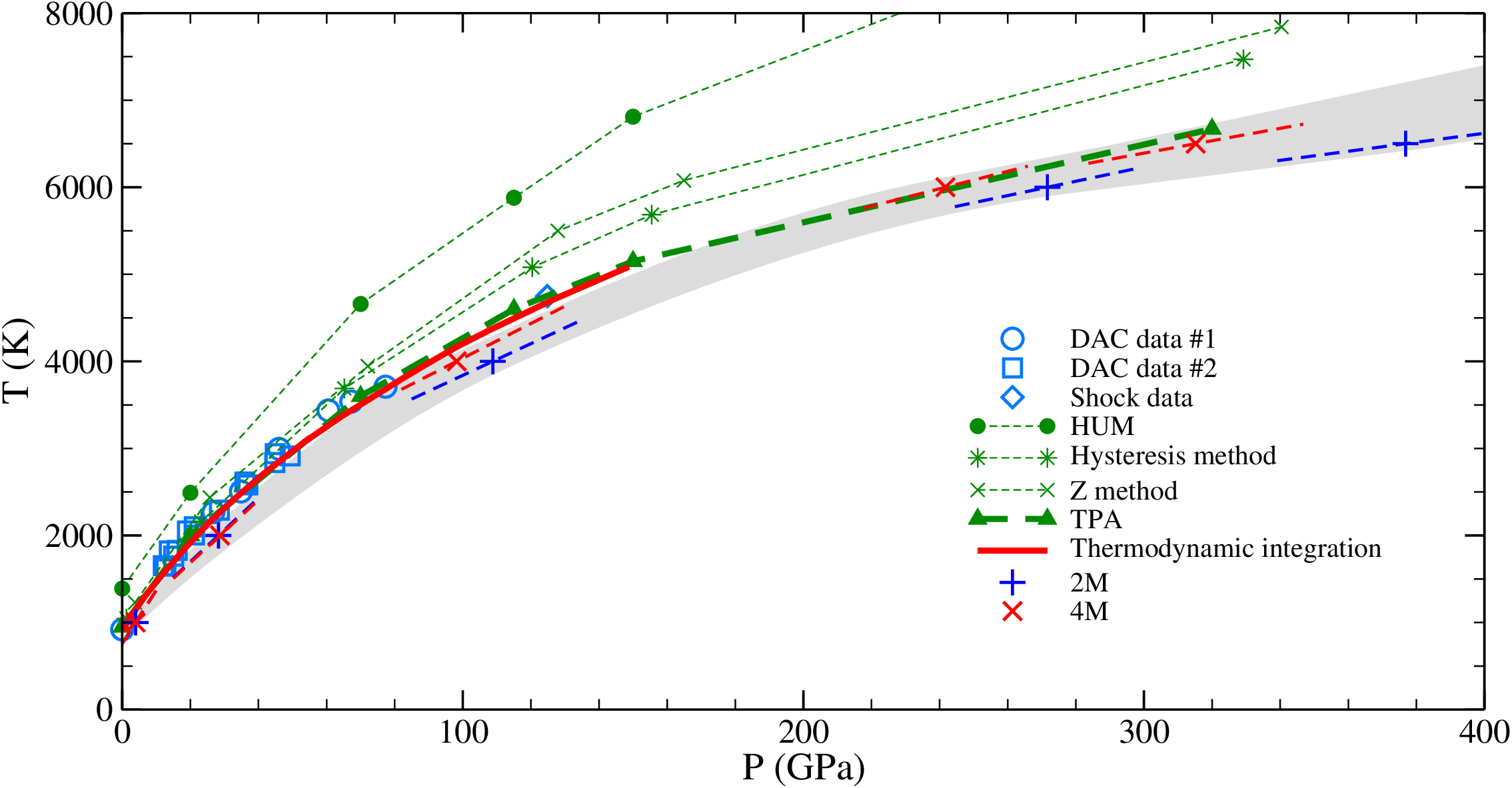}
 \caption{(Color online) Comparison between the present 2PT-MF method and other methods, all within GGA, for the fcc melting curve of aluminum. Experimental data \cite{Shaner1984,Boehler1997,Hanstrom2000} are plotted in open blue symbols. HUM (green filled circles), hysteresis method (green stars), Z method (green crosses) and TPA (green filled triangles) results have been obtained by Bouchet {\it et al.} \cite{bouchet}. Results obtained by Vo{\v c}adlo and Alf\`e \cite{vocadlo} using thermodynamic integration (red solid line) are the reference values. Plus (cross) symbols stand for 2M (4M) method. Dashed bold lines are the derivatives of melting temperatures (Eq. \ref{clapeyron}). Grey area represents the estimated uncertainties.}
 \label{fig:Al_fusion}
 \end{figure*}

For five isotherms at 1000, 2000, 4000, 6000, and 6500\,K, we have determined the melting pressures using the above methodology (Fig.\,\ref{fig:Al_fusion}). 
The simulations were performed with the \textsc{Vasp} code \cite{vasp1,vasp2} in the isokinetic ensemble. The projected augmented wave pseudopotential contains 3 valence electrons and the plane wave cutoff energy is fixed to 300 eV. The exchange and correlation functional is the generalized-gradient approximation (GGA). We have simulated a supercell of 108 atoms, for which the sampling of the Brillouin zone at the $\Gamma$ point is sufficient \cite{bouchet}. The time step was 1\,fs, and the statistical averages were performed on at least 10\,ps. 
This input dataset is representative of a fair compromise between accuracy and computational cost. 
Indeed, we have performed some tests at 4000 K using different cell sizes, k-point values, local density approximation (LDA), time step values as short as 0.2\,fs, without any significant variations of the results. We have also checked that the VACF does not change significantly when evaluated in the microcanonic ensemble instead of the isokinetic ensemble. Estimated uncertainties of the results are displayed further in Figs. \ref{fig:Al_delta} and \ref{fig:Al_fusion} in grey shaded area. Results for T= 4000 K have already been presented in Figs.\,\ref{fig:schema} and \ref{fig:schemab}. Results for the 2M method are given in Table\,\ref{I4000}.

 \begin{table*}
\begin{center}
\begin{tabular}{...........}
\hline
\multicolumn{1}{c}{$V$}  & \multicolumn{1}{c}{~~~~$P_S$~~~~}     & \multicolumn{1}{c}{~~~~$P_L$~~~~} &  \multicolumn{1}{c}{~~~~$e_S$~~~~} & \multicolumn{1}{c}{~~~~$e_L$~~~~} &  \multicolumn{1}{c}{~~~~$s^{i}_S$~~~~} &  \multicolumn{1}{c}{~~~~$s^i_L$~~~~} &  \multicolumn{1}{c}{~~~~$s^e_S$~~~~} &  \multicolumn{1}{c}{~~~~$s^e_L$~~~~} &  \multicolumn{1}{c}{~~~~$G_S$~~~~} &  \multicolumn{1}{c}{~~~~$G_L$~~~~} \\
\multicolumn{1}{c}{(cm$^3$\,g$^{-1}$)} &  \multicolumn{2}{c}{(GPa)} &  \multicolumn{2}{c}{(MJ\,kg$^{-1}$)} & \multicolumn{4}{c}{(kJ\,K$^{-1}$\,kg$^{-1}$)} &  \multicolumn{2}{c}{(MJ\,kg$^{-1}$)} \\
\hline 
0.3333 & 34.66 &  34.62 &  -8.57 &  -8.58 &   3.67 &   3.55 &   0.13 &   0.13  &  -11.68 &  -11.73\\
0.3125 & 44.45 &  44.63 &  -8.12 &  -8.09 &   3.57 &   3.44 &   0.12 &   0.12  &  -8.52 &  -8.40\\
0.2941 & 55.90 &  55.89 &  -7.52 &  -7.52 &   3.47 &   3.36 &   0.12 &   0.12  &  -4.94 &  -5.00\\
0.2778 & 60.99 &  68.65 &  -7.73 &  -6.83 &   3.02 &   3.25 &   0.10 &   0.11  &  -2.88 &  -1.22\\
0.2632 & 74.03 &  82.54 &  -7.11 &  -6.08 &   2.94 &   3.16 &   0.09 &   0.11  &   0.63 &   2.55\\
0.2500 & 89.02 &  98.69 &  -6.34 &  -5.14 &   2.86 &   3.08 &   0.09 &   0.11  &   4.46 &   6.76\\
0.2381 & 105.80 &  115.67 &  -5.46 &  -4.19 &   2.80 &   3.00 &   0.08 &   0.11  &   8.54 &  10.91\\
0.2273 & 124.42 &  134.74 &  -4.46 &  -3.09 &   2.73 &   2.94 &   0.08 &   0.10  &  12.90 &  15.36\\
0.2174 & 144.82 &  150.64 &  -3.35 &  -2.58 &   2.68 &   2.68 &   0.07 &   0.10  &  17.42 &  19.05\\
0.2083 & 167.10 &  169.88 &  -2.14 &  -2.09 &   2.62 &   2.56 &   0.07 &   0.07  &  22.21 &  22.79\\
\hline
\end{tabular}
\end{center}
\caption{Thermodynamic variables along the 4000\,K isotherm of aluminum with the 2M method: volume $V$, pressure $P_S$ ($P_L$), internal energy $e_S$ ($e_L$), ionic and electronic entropies $s^i_S$,$s^e_S$ ($s^i_L$,$s^e_L$), and Gibbs free energy $G_S$ ($G_L$) in the solid fcc (liquid) branch.}
\label{I4000}
\end{table*}

The comparison between the present 2PT-MF method and the other methods, including the reference thermodynamic integration method, is illustrated in Fig.\,\ref{fig:Al_fusion}, together with experimental data \cite{Shaner1984,Boehler1997,Hanstrom2000}. With a low numerical cost, equivalent to twice the HUM cost, 
the 2PT-MF method is almost as accurate as the more expensive TPA method, within 5--10\% of the reference thermodynamic integration method. This is obtained with the 4M option, whereas with the 2M option, a 10--20\% deviation is observed. Such a level of accuracy is to be compared with the differences between melting temperatures measured at low pressures and those obtained within LDA and GGA, of order 5--10\% \cite{vocadlo}.

To further demonstrate the reliability of the 2PT-MF method, we have also compared in Fig.\,\ref{fig:Al_delta} the volume and entropy changes at melting as functions of pressure with the results of the thermodynamic integration approach \cite{vocadlo}. Volume changes are in very good agreement with the reference results of Vo{\v c}adlo and Alf\`e \cite{vocadlo}. For entropy changes, a fair agreement is observed at low pressures. At high pressures, the entropy jump obtained with the 2PT-MF method decreases less with pressure than that observed with thermodynamic integration.

\section{Conclusion}

We have presented a simple, fast, and reliable method to compute the melting curves of materials from {\it ab initio} molecular dynamics simulations. It provides all the needed thermodynamic variables to compare the Gibbs free energies of solid and liquid phases. In particular, the contribution of the nuclei motion to the entropy is calculated in each phase utilizing the two-phase thermodynamic (2PT) model of Lin {\it et al.} \cite{lin}, and its enhanced version given by Desjarlais \cite{desjarlais}, based on the memory function (MF) formalism. In this 2PT-MF method, the only additional quantity to add to the output variables of the simulations is the velocity autocorrelation function. 

To assess the merits of this method with respect to others, we have focused on the fcc melting curve of aluminum, which has been measured at low and medium pressures \cite{Shaner1984,Boehler1997,Hanstrom2000}, and extensively studied theoretically \cite{moriarty,mei,morris,dewijs,jesson,alfe,bouchet}. The 2PT-MF method gives results as accurate as the more elaborate methods,  for a much lower computational cost. The level of accuracy reached in this benchmark is around 10\%, comparable to the discrepancies observed between experiments and {\it ab initio} simulations using different exchange and correlation functionals, LDA and GGA for example. Although more applications of the 2PT-MF method to different chemical elements would be welcome, the present results give very encouraging evidences that it could become the routine calculation of melting curves for many applications. Of special interest would be its use for complex systems such as silica in the cores of exoplanets, gas giants and super-Earth \cite{Mazevet2014}, and for chemical elements with large number of valence electrons, for which other methods would be intractable.

\appendix
\section{Python script for $F_g(\nu)$}

In Eq.\,\eqref{Fnu}, the function $F_g(\nu)$ expresses as:
\begin{equation}
F_g(\nu) = 6 \left[ \frac{1}{\hat K_g(i\xi) + i\xi} + \frac{1}{\hat K_g(-i\xi) - i\xi}\right].\label{eq:A1}
\end{equation}
with $\xi = 2 \pi \nu$ and:
\begin{equation}
\hat K_g(z) = A_g\, \sqrt{\frac{\pi}{4 B_g}}\, e^{z^2/4 B_g}\, \mathtt{Erfc}\left[\frac{z}{2 \sqrt{B_g}}\right].\label{eq:A2}
\end{equation}
It follows from Eq. \ref{eq:A2} and using the notations of \cite{abramowitz} that:
\begin{eqnarray*}
\hat K_g(i\xi) = K_0\, w(-\eta),\\
\hat K_g(-i\xi) = K_0\, w(\eta),
\end{eqnarray*}
with $K_0 = A_g\, \sqrt{\frac{\pi}{4 B_g}}$ and $\eta = \frac{\xi}{2\sqrt{B_g}} = \frac{\pi\nu}{\sqrt{B_g}}$.

Knowing that $w(\overline{z}) = \overline{w(-z)}$ \cite{abramowitz}, we have
\begin{equation*}
\overline{\hat K_g(i\xi)} = \hat K_g(-i\xi).
\end{equation*}
Eq. \ref{eq:A1} then reduces to:
\begin{equation*}
F_g(\nu) = 6 \left[ \frac{\hat K_g(i\xi) + \overline{\hat K_g(i\xi)} }{(\hat K_g(i\xi) + i\xi)\cdot(\overline{\hat K_g(i\xi) + i\xi})} \right]
\end{equation*}
One can see that the function $F_g(\nu)$ returns a real value. Moreover the function $w(z)$ can be easily computed with a {\tt Python} script using the {\tt Scipy} package ({\tt scipy.special.wofz(z)}). The function $F_g(\nu)$ is then obtained by:

\begin{verbatim}

import numpy as np
import scipy.constants as sc
from scipy.special import wofz
def function_Fg(nu,Ag,Bg):
    K0  = Ag * np.sqrt(sc.pi/Bg) / 2.
    eta = sc.pi * nu / np.sqrt(Bg)
    Kg    = K0 *  wofz( -eta )
    num   = Kg + np.conj(Kg)
    Kg   += complex( 0., 2. * sc.pi * nu )
    denom = Kg * np.conj(Kg)
    Fg    = 6. * num / denom
    return Fg
	
\end{verbatim}


\end{document}